# Aligned Graphene Nanoribbons and Crossbars from Unzipped Carbon Nanotubes

Liying Jiao[1], Li Zhang[1], Lei Ding[2], Jie Liu[2] & Hongjie Dai[1]*

1. Department of Chemistry and Laboratory for Advanced Materials, Stanford University, Stanford, California 94305, USA

2. Department of Chemistry, Duke University, Durham, North Carolina 27708, USA

*Correspondence to hdai@stanford.edu*

**Abstract**

**Aligned graphene nanoribbon (GNR) arrays were made by unzipping of aligned single-walled and few-walled carbon nanotube (CNT) arrays. Nanotube unzipping was achieved by a polymer-protected Ar plasma etching method, and the resulting nanoribbon array was transferred onto any substrates. Atomic force microscope (AFM) imaging and Raman mapping on the same CNTs before and after unzipping confirmed that ~80% of CNTs were opened up to form single layer sub-10 nm GNRs. Electrical devices made from the GNRs (after annealing in $H_2$ at high temperature) showed on/off current ($I_{on}/I_{off}$) ratios up to $10^3$ at room temperature, suggesting semiconducting nature of the narrow GNRs. Novel GNR-GNR and GNR-CNT crossbars were fabricated by transferring GNR arrays across GNR and CNT arrays, respectively. The production of ordered graphene nanoribbon architectures may allow for large scale integration of GNRs into nanoelectronics or optoelectronics.**

Graphene nanoribbons have emerged as an interesting material with a wealth of electronic and spin transport properties [1-6]. Narrow GNRs (sub-10 nm) open up band gaps useful for field effect transistors (FETs) [1-4]. Several approaches have been developed to make GNRs, such as lithographic patterning [5-9], sonochemical methods [1, 10], chemical vapor deposition (CVD) [11, 12], nanocutting [13-15] and unzipping of carbon nanotubes (CNTs) [16-22]. In addition to controlling the width, edge smoothness and quality of GNRs, it is important to control the alignment of GNRs for their integration into devices. Among existing methods, lithographic patterning is capable of fabricating large arrays of aligned GNRs but the width and edge smoothness of GNRs are not well controlled, especially at the sub-10 nm scale. Unzipping CNTs offers a possibility to meet these challenges since much progress has been made on controlled synthesis (alignment, density, length and diameter controls) of CNTs in the past two decades [23-25]. Our pervious work demonstrated that GNRs with well controlled width in the 10-20 nm range and smooth edges could be made by unzipping multiwalled carbon nanotubes (MWNTs) using a polymer-protected plasma etching method [16]. Most recently, we developed a new method that produced highly pristine nanoribbons by unzipping mildly gas-phase oxidized MWNTs sonochemically in an organic solvent [22]. Thus far, GNRs from unzipped CNTs have lacked alignment and ordering on substrates required for device integration.

Here, we show that highly aligned, narrow GNRs can be made from aligned array of single-walled carbon nanotubes (SWNTs) by our polymer-protected plasma etching method (Fig. 1 (a) and (b)). Aligned CNTs were first partially embedded in a polymer film and the unprotected sides of CNTs were then exposed to Ar plasma. The polymer film served as an etching mask and enabled the longitudinal unzipping of CNTs [16]. The alignment of CNTs was maintained through embedding in the polymer film, unzipping by plasma and transferring to another substrate. Considering the topographic similarity between narrow GNRs and SWNTs, we performed careful atomic force microscope (AFM) imaging and Raman spectroscopic measurements on the same CNTs before and after the unzipping process to obtain spectroscopy evidence of successful unzipping of CNTs into GNRs. Both the obvious decrease in topographic height and changes in Raman spectra of the same CNTs after unzipping confirmed the creation of GNRs. Field-effect transistors (FETs) with individual and array of GNRs as channel materials showed on/off current ($I_{on}/I_{off}$) ratios up to $10^3$ and 20, respectively. Our method also allowed for the formation of crossbars of GNR-GNR and GNR-CNT for the first time.

The starting materials used were dense (1~2 tubes/μm), uniform, highly aligned arrays of long (~1 millimeter) CNTs (diameter: 0.6-3.4 nm; ~80% were SWNTs and ~20% were few-walled CNTs and bundles) grown on ST-cut quartz by CVD [26]. We used makers to track specific CNTs through the unzipping process. We first transferred as-grown CNT array from quartz to $SiO_2$/Si substrates with sputtered gold maker array by a nano-transfer printing technique [27]. After that, the CNT array was unzipped following a similar procedure as our previous work [16]. Briefly, a thin film of poly (methyl methacrylate) (PMMA) was spin-coated on top of the CNT array. After baking, CNTs and gold markers were embedded in the PMMA film. The PMMA/CNTs film with gold markers was then peeled off in a KOH solution. The film was then exposed to 5 W Ar plasma for various times. With the protection of PMMA, only the exposed part of tube sidewall was etched by the Ar plasma. After etching, the film was contact-printed to another SiO$_2$/Si substrate. Finally, PMMA film was removed by acetone vapor and followed by calcination at 300 $^o$C for 20 min, leaving the unzipped GNRs and gold markers on the target substrate. The protection of PMMA afforded longitudinal unzipping of CNTs and maintained the alignment of CNTs through the whole process, both were critical to the creation of highly aligned GNR arrays.

AFM characterization found that the resulting GNRs were mostly single-layered with average height of ~1.0 nm (Fig. 1(d) and Fig. S-1(b) in Electronic Supplementary Material (ESM)).The starting CNTs exhibited an average diameter of ~1.4 nm (Fig. 1(c) and Fig. S-1(a) in ESM). The GNR array maintained the main features of their parent CNT array including high density, good alignment and ultra-long length. The widths of most obtained GNRs fall into the range of 2-8 nm (see ESM, Fig. S-1 (c)). As shown in the AFM images (Fig. 1 (e)-(g) and Fig. S-2 in ESM), the edges of the GNRs appeared smooth and uniform along the length of GNRs. GNR-CNT junction-like structures were also observed in some partially unzipped tubes due to nonuniform etching caused by local deformations of the PMMA film (Fig. 1(h)). The etching condition was critical to the successful unzipping of SWNTs. We found that the optimized etching condition was 5 W (the lowest stable power we could use) for 10 s, which was able to unzip the majority of tubes without introducing many breaks.

By locating specific CNTs with makers, we obtained AFM images and Raman spectra of the same CNTs before and after plasma etching. As shown in Fig. 2 (a) and (b), the height of CNTs decreased by ~0.4-0.9 nm after the unzipping process, which indicated that the CNTs were opened up into GNRs by the Ar plasma. We found that the unzipping efficiency was dependent on the diameter of parent tubes. Small SWNTs (diameter <1.0 nm) were totally etched by plasma and CNTs with a diameter of > 2.5 nm were not unzipped by the condition used. All the CNTs with medium diameters (1.0-2.5 nm) were successfully converted into sub-10 nm GNRs. Therefore, the yield of GNRs (~80 %) was limited by the diameter distribution of the starting CNTs. Besides the decrease in height, important changes were also observed in Raman spectra of CNTs after unzipping. The Raman spectra of pristine CNTs showed strong G-band and no obvious D-band, which indicated high quality of the starting materials [28]. After the unzipping process, the Raman G-band intensity decreased by ~60% and strong D-band appeared (Fig. 2(c) to (f)). The appearance of D-band was attributed to the creation of edges that acted as defects responsible for momentum conservation in the double resonance Raman process [29, 30]. The D-band to G-band ($I_D/I_G$) ratio of the GNRs in the array was ~1.5, much higher than their parent raw tubes (<0.01). The obvious increase of $I_D/I_G$ ratio also confirmed that the CNTs were unzipped by our method. We also performed Raman mapping over individual GNRs (see ESM, Fig. S-3) and the averaged $I_D/I_G$ ratio of individual GNRs was ~1.5. The averaged $I_D/I_G$ ratio of the sub-10 nm GNRs obtained here was higher than wider GNRs made by unzipping MWNTs by the same approach due to much narrower width (2-8 nm) and proportionally higher defects density at the edges [16].

We fabricated FETs on individual and arrays of GNRs, with palladium (Pd) as source/drain (S/D) metal contacts (channel length $L$ ~ 100 nm), a p++ Si backgate, and 100 nm SiO$_2$ as gate dielectrics. The resulting devices of as-made single GNRs showed higher resistance and lower $I_{on}/I_{off}$ ratio (<$10^2$) at room temperature than

sub-10 nm GNRs made by the previous sonochemical approach [1], which suggested high defect density in the as-made GNRs. To improve quality, we carried out thermal annealing aimed at reducing defects on the GNRs generated by the plasma etching process. We found that thermal annealing in 1 Torr of $H_2$ at 800 $^o$C for 20 min improved the electrical properties of GNRs. The $I_{on}/I_{off}$ ratios and on-state conductance improved. Figure 3(a) and (b) show a ~2 nm wide annealed GNR with an $I_{on}/I_{off}$ ratio of ~500 and on-state current of ~500 nA at a bias voltage ($V_{ds}$) of -1V. We also fabricated FET devices on a small array of aligned GNRs. A FET made of 3 aligned GNRs showed an $I_{on}/I_{off}$ ratio of 20 and on-state current of 1.5 μA at $V_{ds}$= -1 V (Fig. 3(c) and (d)). These results suggested that the sub-10 nm GNRs made by the PMMA-protected plasma unzipping method contained a high defect density, partly due to the ultra-narrowed widths of the ribbons. Strategies should be devised to improve the quality of aligned narrow GNRs for practical application in nanoelectronics.

The aligned GNRs were explored as building blocks for constructing complex two dimensional (2D) structures. We fabricated crossbar structures potentially useful as logic and memory elements in nanoelectronics [31-33] using GNR arrays. Crossbar arrays of GNRs were fabricated by transferring aligned GNR array on top of another array with a rotation of 90$^o$. The two layer of GNR arrays formed well ordered square mesh structures as evidenced by AFM and Raman G-band images (Fig. 4(a) and (b)). Figure 4 (c) shows an AFM image of a GNR-GNR junction. Besides GNR-GNR crossbars, we also made two configurations of GNR-CNT crossbars array by transferring GNRs on top of CNTs or CNTs on top of GNRs (Fig. 4 (d) and Fig. S-4 (a) in ESM). In the GNR-CNT junction, a ~6 nm wide single layer GNR was placed on the top of a CNT (diameter: ~3.3 nm) and introduced a radial deformation of ~0.3 nm to the underneath CNT at the cross point (Fig. 4(d) and Fig. S-4 (c) in ESM), which was also observed in CNT-CNT junctions [34] (Fig. S-4(b) and (d) in ESM). No obvious deformation was found when the CNTs were on top of GNRs (Fig. S-4 (a) in ESM). The successful fabrication of GNR-GNR and GNR-CNT crossbars will make it possible to explore the fundamental properties and possible applications of these junctions.

In summary, well aligned narrow (<10 nm) GNR arrays have been made from CNT arrays by polymer-protected plasma unzipping. Well ordered 2D architectures of GNR-GNR and GNR-CNT crossbars were constructed. Our approach is compatible with semiconductor processing to obtain GNR arrays. However, future work is required to improve the quality of the narrow GNRs unzipped from small nanotubes for applications in high performance nanoelectronics. Nevertheless, the current work represents the beginning of making GNR arrays from nanotube arrays and opens up a way to large scale integration of GNRs.

**Experimental**

**Preparation of GNRs array**

Aligned CNTs were transferred from quartz substrate to $SiO_2$/Si substrate with prefabricated marker array (formed by EBL and sputtering of 30 nm gold) using the procedures described in ref 27. Selected CNTs were located with the aid of markers and characterized using AFM and Raman mapping. After that, a PMMA solution ($M_w$ = 495 K, 5% in anisole) was spin-coated on the substrate at 3000 round per minute (r.p.m) for 1 min and then baked at 170 °C for 2 hrs on a hot plate. The PMMA film was peeled off together with CNTs and gold markers in 1M KOH solution at 80 °C. Then the film was rinsed with water and dried in air. After that, the film was exposed to 5 W Ar plasma for 10 s in a plasma reactive ion etching (RIE) system (MRC Model 55) at a base pressure of 40 mTorr. After etching, the film was printed onto a $SiO_2$/Si substrate. Then PMMA was removed with the vapor of acetone, leaving the unzipped CNTs and gold markers on the target substrate. Finally, the substrate was calcined at 300 °C for 20 min to remove PMMA residue. To improve the quality of GNRs, the calcined GNRs were annealed in $H_2$ at 800 °C for 20 min at a pressure of 1 Torr.

**Characterization of GNRs**

AFM images of the GNR were taken with a Nanoscope IIIa multimode instrument in tapping mode. Raman spectra of GNRs were collected with Horiba Jobin Yvon LabRAM HR Raman microscope with a 633 nm He-Ne laser excitation (spot size ~1 μm, power ~10 mW). The step size of mapping was 100 nm and the integration time was 10 s at each spot.

**Device fabrication**

Before making devices, we first patterned the dense and long GNR arrays to within isolated regions using EBL and $O_2$ plasma etching. We then did a second EBL followed by electron beam deposition of Pd (30 nm) and lift-off to fabricate arrays of source- and drain-electrodes on the GNRs. The channel length of these devices was ~200 nm and the width of source and drain electrodes varied from 500 nm and 5 μm for single- and multiple ribbon devices, respectively. The devices were annealed in Ar at 220 ºC for 15 min to improve the contact quality. AFM was then used to identify devices with a single or multiple GNR connection. Electrical characterization of the devices was carried out in air using a semiconductor analyzer (Agilent 4156C).

**Acknowledgments**

This work was supported by MARCO-MSD, Intel, ONR and graphene-MURI.

**Electronic Supplementary Material:** Further characterization of aligned GNR array and crossbars of GNR-CNT and CNT-CNT by AFM and Raman can be found in the ESM with four figures which are available in the online version of this article.

**Legend**

**Figure 1. GNR array made from CNT array.** (a) and (b), Schematic of making GNR array from CNT array by PMMA-protected plasma etching. (c) and (d), Typical AFM images of pristine CNT array and obtained GNR array, respectively. Insets in (c) and (d) are zoom-in images of individual CNT and GNR, respectively. (e)-(g), AFM images of individual GNRs with different widths. The widths for the GNRs from (e) to (g) were 8 nm, 4 nm and 2 nm, respectively. (h) AFM image and schematic of CNT-GNR junctions in a partially unzipped tube.

**Figure 2. Characterization of the same CNTs before and after unzipping.** (a) and (b), AFM images of the same CNT array before and after unzipping process. Insets: zoom-in AFM images of the same CNTs before and after unzipping. The height decrease of these three tubes shown in (a) after unzipping were 0.7 nm, 0.4 nm and 0.9 nm from left to right. (c) and (d), Raman G-band images of the same CNTs shown in (a) and (b), respectively. The intensity of G-band decreased significantly after the unzipping process. (e) and (f), Averaged Raman spectra of the same CNTs shown in (a) and (b), respectively.

**Figure 3. FETs of individual and a small array of GNRs.** (a) Source-drain current versus gate voltage ($I_{ds}$-$V_{gs}$) curve of a ~2 nm wide single layer GNR probed in air at $V_{ds}$=10 mV. Inset, AFM image of the GNR-FET. The $I_{on}$/$I_{off}$ ratio of this GNR device was ~500. (b) Current-voltage ($I_{ds}$-$V_{ds}$) curves for the device in (a) at various gate biases $V_{gs}$ from -30 V to 30 V at a step of 5 V from bottom to top. (c) $I_{ds}$-$V_{gs}$ curve of a FET made on an array of 3 GNRs at $V_{ds}$=10 mV. Inset: AFM image of the GNR-FET. The widths of these GNRs were ~5 nm, ~3 nm and ~4 nm from left to right. The $I_{on}$/$I_{off}$ ratio of this device was ~20. (d) $I_{ds}$-$V_{ds}$ curves for the device in (c) at various gate biases $V_{gs}$ from -30 V to 30 V at a step of 5 V from bottom to top.

**Figure 4. Crossbars of GNR-GNR and GNR-CNT.** (a) AFM image of crossbar array of GNRs. (b) Raman G-band image of the same crossbar array shown in (a). (c) 3D AFM image and schematics of GNR-GNR crossbar. The widths of both GNRs were ~5 nm. (d) 3D AFM image and schematics of GNR-CNT crossbar. The ~6 nm wide GNR was on the top of a CNT with a diameter of ~3.3 nm. The height of the CNT near the cross point decreased to 3.0 nm due to the radial deformation introduced by the GNR (Fig. 4-(c) in ESM).

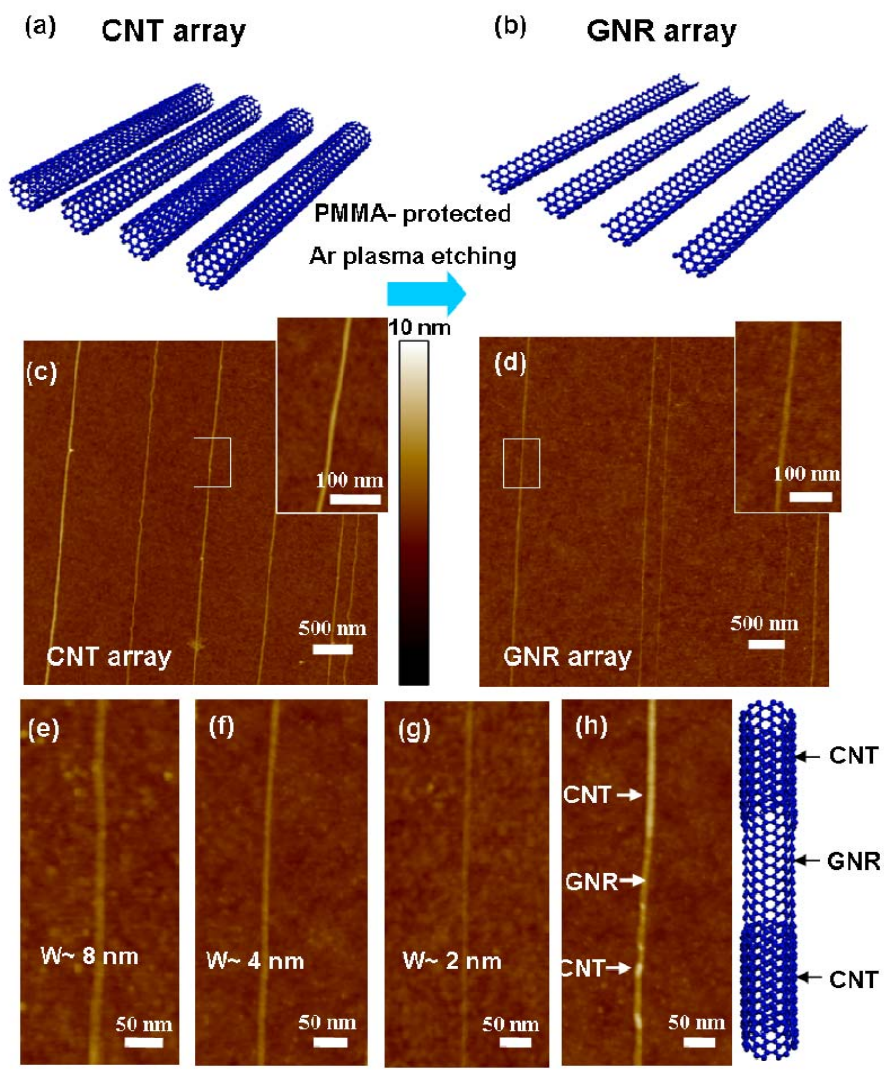

**Figure 1**

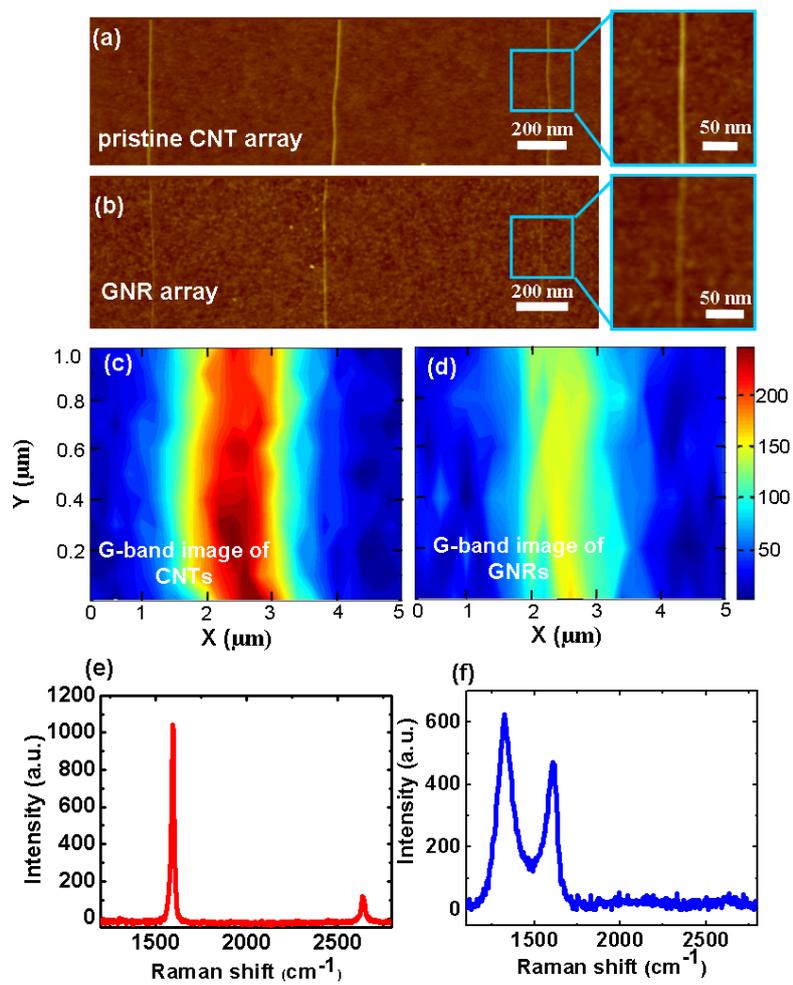

**Figure 2**

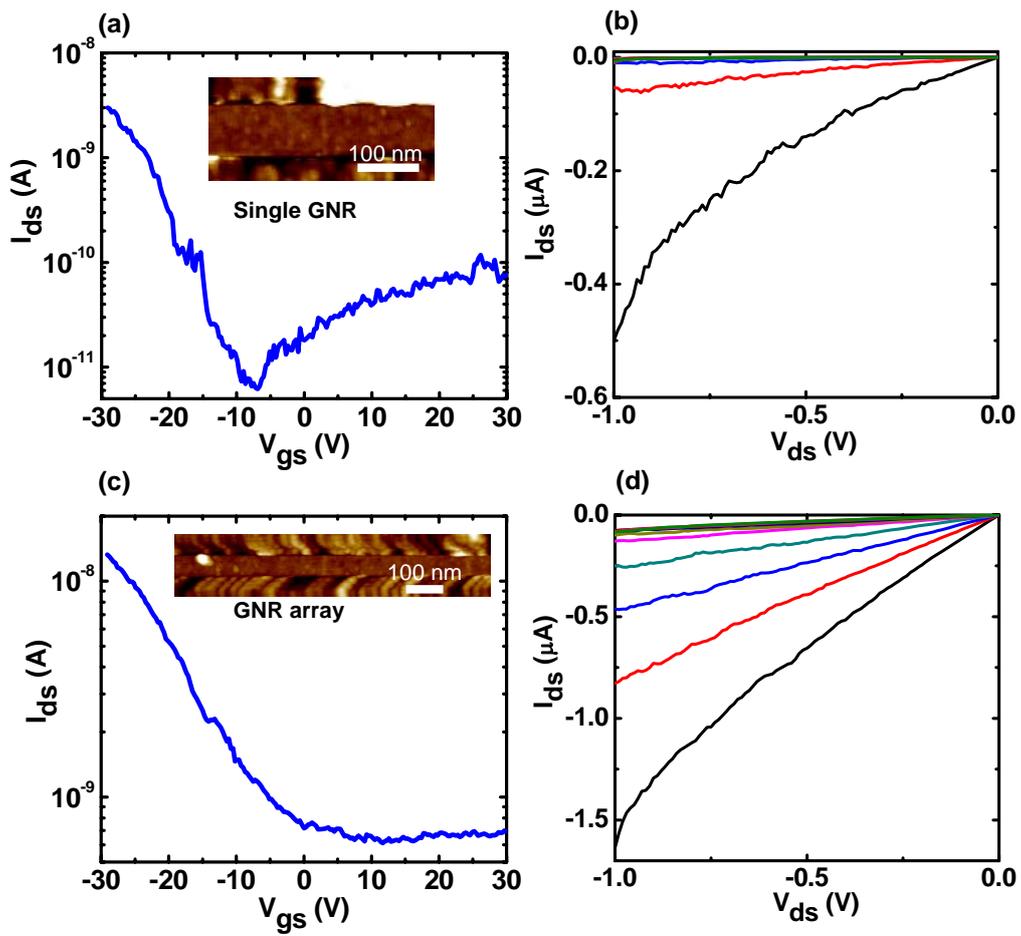

**Figure 3**

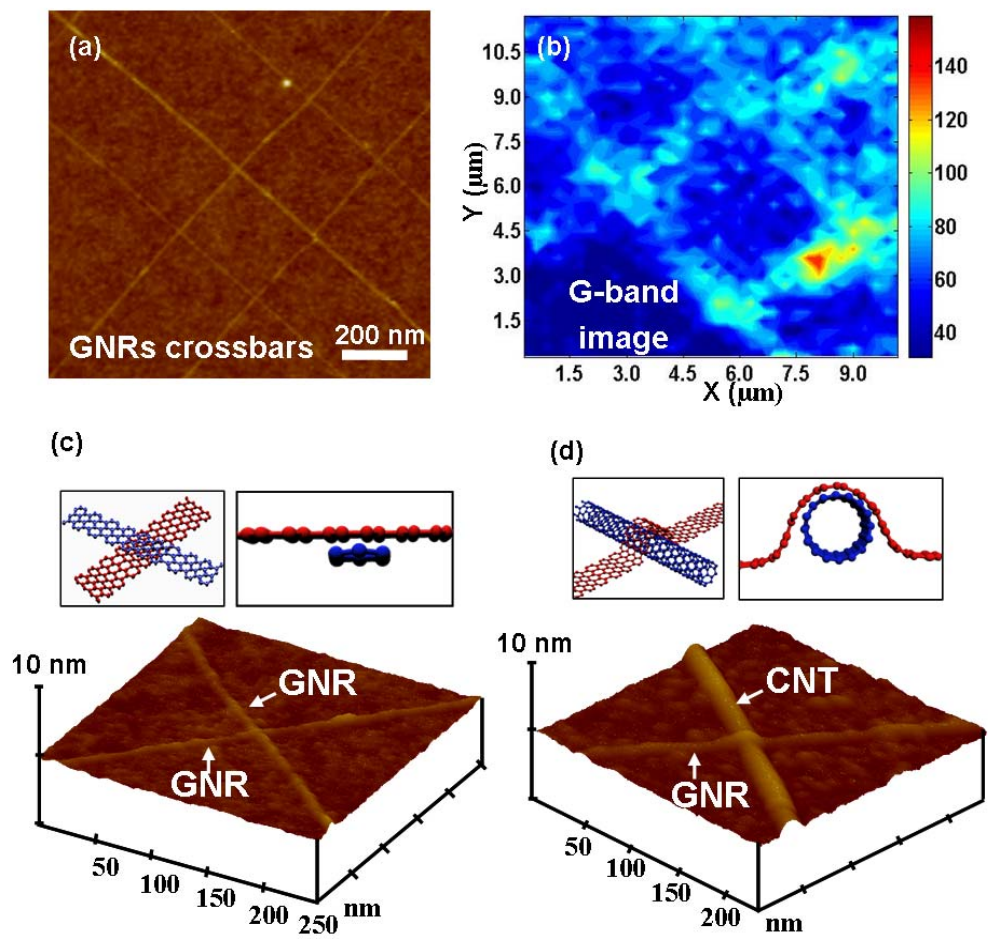

**Figure 4**